\documentclass[aps,prl,reprint]{revtex4-1}
\usepackage{graphicx}
\usepackage{dcolumn}
\usepackage{bm}
\usepackage{hyperref}
\usepackage{amssymb}
\usepackage{amsmath}
\usepackage{epsf}
\usepackage[caption=false]{subfig}
\usepackage{epstopdf}
\usepackage{amsfonts}
\usepackage{color}

\usepackage{pst-all}

\begin{document}
\title{Statistics of an adiabatic charge pump}
\author{Hari Kumar Yadalam and Upendra Harbola}
\affiliation{Department of Inorganic and Physical Chemistry, Indian Institute
of Science, Bangalore, 560012, India.}

\begin{abstract}
We investigate the effect of time-dependent cyclic-adiabatic driving on the
charge transport in quantum junction. We propose a nonequilibrium Greens 
function formalism to study statistics of the charge pumped (at zero bias) through the junction. 
The formulation is used to demonstrate charge pumping in
a single electronic level coupled to two (electronic) reservoirs with
time dependent couplings. Analytical expression for the average pumped 
current for a general cyclic driving is derived. 
It is found that for zero bias, for a certain class of
driving, the Berry phase contributes only to the odd cumulants.
To contrast, a quantum master equation formulation 
does not show Berry-phase effect at all. 
\end{abstract}
\maketitle

\begin{figure}[h]
\centering
\includegraphics[width=8cm]{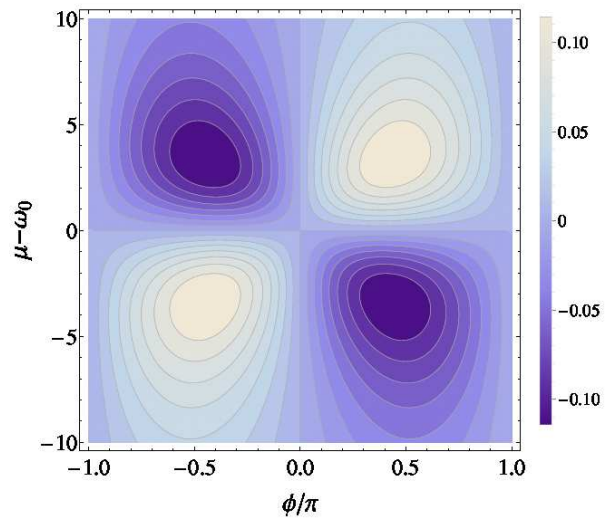}
\caption{Average number of particles pumped per cycle, $N_{pump}$ as a function
of $\mu-\omega_0$ and $\phi$. Here $\Gamma_L(t)=1+10 \sin^2(\frac{\pi
t}{T_p}-\frac{\phi}{4})$, $\Gamma_R(t)=1+10 \sin^2(\frac{\pi
t}{T_p}+\frac{\phi}{4})$. Parameters $\Gamma_\alpha, \mu$ and $\omega_0$ are 
in units of $\beta^{-1}$. }
\label{fig-1}
\end{figure}

It is well known that the effect of adiabatically  varying few parameters in the
Hamiltonian in a cyclic manner enters in the wavefunction in the
form of a phase factor. This phase factor consists of two parts, one is called
dynamical part (which, in general, depends on how fast the parameters are varied)
and the second one, generally known as Berry phase (also called geometric phase)
that depends only on the path (area) traced in the parameter space and
independent of how fast it is traced provided adiabatic condition is
satisfied\cite{Berry1984,Chruscinski2012,Bohm2013}.
Somewhat counter-intuitive, this phase factor may lead to changes in macroscopic
observables, like finite spin or charge currents in one dimensional phase
coherent rings at equilibrium\cite{Cheung1988,Balatski1990,Splettstoesser2003}.
Originally developed in the context of closed quantum systems, recent works have
extended the geometric phase concept to the case of open quantum systems out of
equilibrium\cite{Zhou2003,Albash2012}. This is usually treated
within the quantum master equation (QME) approach\cite{Harbola2006}.  Stochastic
variation of system parameters is known to induce net flux in open systems like
quantum heat pumps\cite{Ren2010,Segal2008}, quantum electron
pumps\cite{Yuge2012} and also classical stochastic systems like enzyme kinetics,
molecular motors and living cell locomotion\cite{Sinitsyn2009}.
On a similar footing, adiabatic cyclic variation of parameters in the
Hamiltonian may also lead to finite flux \cite{Brouwer1998}.
Switkes et.al.,\cite{Switkes1999}  have experimentally demonstrated an adiabatic
quantum pump by modulating confining potential of an open quantum dot in a
cyclic manner, leading to a finite voltage drop across the quantum dot.
Modifying the potential at two ends of the dot changes the character of the
wavefunction and therefore modifies the couplings to the electron reservoirs. In
this work we explore this aspect within the most general framework based on 
non-equilibrium Greens function \cite{Kita2010}.
 The adiabatic driving may also effect the statistics of charge transfer and the
steady-state fluctuation relation due to Gallavoti Cohen (GC) type
symmetry \cite{Esposito2009} may also get modified.
Ren et.al. \cite{Ren2010} have recently used QME to study heat pumping and
fluctuations of heat transfer in a two-level system sandwiched between two
thermal reservoirs. It was shown that in the case of time-dependent temperature
modulations of the two heat reservoirs, heat transfer statistics does not admit
GC type symmetry.
It was also argued that modulating couplings to thermal reservoirs does not lead
to any pumping.
Several methods like scattering theory\cite{Andreev2000,Levitov2001}, Floquet
scattering theory\cite{Moskalets2004}, adiabatic master equation
approach\cite{Albash2012,Yuge2012}  etc., have been developed for studying the
statistics of adiabatic pumping.
But each of these methods has its own advantages and short-comings. 
In this work we develop a scheme within NEGF formalism (which in
principle is exact) and apply it to study the effect of cyclic-adiabatic driving
on the charge transfer statistics in a resonant level model.
We find that finite charge transfer between two reservoirs (at the same
thermodynamic states) connected through a single level quantum system is
possible by modulating the couplings to the reservoirs in an adiabatic-cyclic
manner. We present an analytic expression for the pumped current for a general driving. 
The direction of the net charge flow can be varied by changing the sign of the phase difference
between the two time-dependent couplings. We find that the flux direction also
depends on the energy difference between the chemical potentials ($\mu$) of reservoirs
and the level energy ($\mu-\omega_0$). In general, the charge transfer fluctuations 
are modified due to the Berry phase. However, at equilibrium, for certain class of drivings,
asymmetric fluctuations (odd cumulants) are generated solely due to the Berry-phase.   
The full statistics of the pumped charge satisfy 
a steady-state fluctuation relation. 
We emphasize that the present formulation shows that
it is possible to pump a finite net charge in {\em noninteracting} open quantum
junctions, unlike a simple QME formulation which does not lead to any pumping due to the
Berry-phase \cite{Yuge2012,Ren2010}.\\
{\bf Model Hamiltonian:}
A general Hamiltonian for the description of electron transport in a quantum
junction where a molecular system is coupled to two (non interacting) electronic
reservoirs is
\begin{eqnarray}
\label{eq-1}
 \hat{H}(t)&=&\sum_r\epsilon_r(t)^{} d_{r}^\dag d_{r}^{}+H_{int}+\sum_{\alpha
k}\epsilon_{\alpha,k}^{}c_{\alpha k}^\dag c_{\alpha k}^{}\nonumber\\
         &+&\sum_{r,\alpha k}\big[g_{\alpha k,r}^{*}(t)c_{\alpha k}^\dag
d_{r}^{}+g_{\alpha k,r}^{}(t)d_{r}^\dag c_{\alpha k}^{} \big]
\end{eqnarray}
where $d_r^\dag$($d_r$) stands for electron creation (annihilation) operator in
the $r$-th system orbital while $c^\dag_{\alpha,k}$ ($c_{\alpha,k}$) is for 
electron 
creation (annihilation) operator on the left or right ($\alpha=L/R$) reservoir in
the energy  state $\epsilon_{\alpha,k}$. $H_{int}$ is the Hamiltonian to account
for all other possible interactions in the system, like Coulomb and
electron-phonon interactions. Here system lead couplings and/or single electron
orbital energies can be periodically modulated (which can be experimentally
realized by applying time dependent gate voltages).
In this work we consider the case when the driving time period is large compared
to the internal relaxation time scales in the system such that the system at any
time is at steady state with respect to reservoirs.\\
{\bf Two-point measurement:}
Under these conditions, we consider two simultaneous measurements of electron number
in the left and the right reservoirs (as $[N_L,N_R]=0$, simultaneous measurement
is quantum mechanically allowed) at time $T_0$ and $T>T_0$. 
The probability of change in the number of
particles, $N_{\alpha}(T)-N_{\alpha}(T_0)$, in the left and the right reservoirs
during the measurement time, $T-T_0$, can be computed using the generating
function (GF) containing corresponding counting parameters $\lambda_L$ and
$\lambda_R$ as
\begin{eqnarray}
\label{eq-2}
P(n_L,n_R,T-T_0)&&=\int_0^{2\pi}\frac{d\lambda_L}{2\pi}\int_0^{2\pi}\frac{d\lambda_R}{2\pi}\nonumber\\
 &&{\cal Z}(\lambda_L,\lambda_R,T-T_0) e^{i(\lambda_L n_L+\lambda_R n_R)}
\end{eqnarray}
Using the two time quantum measurement formalism the generating function for
particle number counting on both reservoirs can be written as
\cite{Esposito2009}
\begin{eqnarray}
\label{eq-3}
 &&{\cal Z}(\lambda,T-T_0)=\nonumber\\
 &&\big\langle e^{-i(\lambda_L N_L(T)+\lambda_R N_R(T))} e^{i(\lambda_L
N_L(T_0)+\lambda_R N_R(T_0))} \big\rangle_{\rho(T_0)}
\end{eqnarray}
where $\rho(T_0)$ is the density matrix of the system + reservoirs at time $T_0$
and $N_L=\sum_kc_{L k}^\dag c_{L k}^{}$, $N_R=\sum_kc_{R k}^\dag c_{R k}^{}$ are
the particle number operators corresponding to left and right reservoirs
respectively.
Assuming $[\rho(T_0),N_L]=0$ and $[\rho(T_0),N_R]=0$ (or, more generally, system
and reservoirs are decoupled at time $T_0$ and each are at equilibrium
independently), Eq.(\ref{eq-3}) can be recast as
\begin{align}
\label{eq-4}
 {\cal Z}&(\lambda_L,\lambda_R,T-T_0)\nonumber\\
 &=\bigg\langle {\cal U}_{(\frac{\lambda_L}{2},\frac{\lambda_R}{2})}(T_0,T)
{\cal
U}_{(-\frac{\lambda_L}{2},-\frac{\lambda_R}{2})}(T,T_0)\bigg\rangle_{\rho(T_0)}
\nonumber\\
 &=\bigg\langle{\cal T}_c e^{-\frac{i}{\hbar}\int_c
H^{\lambda(\tau)}_T(\tau)d\tau}\bigg\rangle_{\rho(T_0)}
\end{align}
where in the last line time dependent $\lambda(\tau)$ has been defined on the
Keldysh contour\cite{Rammer2007,Haug2008} (which goes from $T_0$ to $T$ and back
to $T_0$)\cite{Gogolin2006} as
$\lambda(\tau)=(-\frac{\lambda_L}{2},-\frac{\lambda_R}{2})$ on the forward
contour and 
$\lambda(\tau)=(\frac{\lambda_L}{2},\frac{\lambda_R}{2})$ on the backward
contour. Where $\mathcal{T}_c$ refers to time ordering operator on the Keldysh
contour. The evolution on the Keldysh contour is with respect to $\lambda(\tau)$
dependent Hamiltonian (note that $\lambda$-dependent evolution is no longer
unitary), and evolves the ket and the bra with different $\lambda$-dependent
Hamiltonian \cite{Bagrets2003}, which can be obtained by replacing the last line
in Eq. (\ref{eq-1}) by
$\sum_{r,\alpha k}\big[g_{\alpha k,r}^{*}(\tau)e^{i\lambda_\alpha
(\tau)}c_{\alpha k}^\dag d_{r}^{}+g_{\alpha k,r}^{}(\tau)e^{-i\lambda_\alpha
(\tau)}d_{r}^\dag c_{\alpha k}^{} \big]$.
Hence the effect of measurement is reflected in the form of modified
($\lambda$-dependent) couplings to the reservoirs\cite{Shelankov2003}.\\
{\bf GF in terms of NEGF:}
Taking $\lambda_R$ derivative \cite{Gogolin2006} of the logarithm of
Eq.(\ref{eq-4}), we get 
\begin{eqnarray}
\label{eq-6}
 &&\bigg[\frac{\partial\mbox{ln}{\cal
Z}(\lambda_L,\lambda_R,T-T_0)}{\partial(i\lambda_R)}\bigg]=\frac{1}{2}\sum_{k,r}
\int_{T_0}^T dt_1\nonumber\\
 &&
\big[g_{Rk,r}^{*}(t_1)e^{-i\frac{\lambda_R}{2}}G_{r,Rk}^{++}(t_1,t_1)-g_{Rk,r}^{
}(t_1)e^{i\frac{\lambda_R}{2}}G_{Rk,r}^{++}(t_1,t_1)\nonumber\\
 &&+
g_{Rk,r}^{*}(t_1)e^{i\frac{\lambda_R}{2}}G_{r,Rk}^{--}(t_1,t_1)-g_{Rk,r}^{}
(t_1)e^{-i\frac{\lambda_R}{2}}G_{Rk,r}^{--}(t_1,t_1) \big]\nonumber\\
\end{eqnarray}
where $G_{Rk,r}^{++}(t,t')$, $G_{r,Rk}^{++}(t,t')$, $G_{Rk,r}^{--}(t,t')$ and
$G_{r,Rk}^{--}(t,t')$ are appropriate real time projections of mixed contour
ordered Greens functions between system orbitals and reservoir states defined as
$G^{c}_{Rk,r}(\tau,\tau')=-\frac{i}{\hbar}\langle\mathcal{T}_c c_{Rk}(\tau)
d_r^\dag(\tau')\rangle$ and
$G^{c}_{r,Rk}(\tau,\tau')=-\frac{i}{\hbar}\langle\mathcal{T}_c d_r(\tau)
c_{Rk}^\dag(\tau')\rangle$.
Here '+' ('-') index refers to the time variable located on the upper (lower)
Keldysh contour\cite{Rammer2007}. Equation (\ref{eq-6}) can be recast in terms
of the system Greens function matrix alone \cite{Haug2008} which can be
expressed in terms of Wigner transformed
quantities \cite{Kita2010}. In the large measurement time limit, Eq. (\ref{eq-6}) can 
be recast as \cite{sup},
\begin{eqnarray}
\label{eq-7}
&&\bigg[\frac{\partial\mbox{ln}{\cal
Z}(\lambda_L,\lambda_R,T-T_0)}{\partial(i\lambda_R)}\bigg]=\int_{T_0}^{T}dt\int_
{-\infty}^{+\infty}\frac{d\omega}{2\pi}\nonumber\\
&&\mbox{Tr}\bigg[\breve{\Sigma}_R^{+-}(\omega,t)\mathbb{\breve{G}}^{-+}(\omega,
t)-\mathbb{\breve{G}}^{+-}(\omega,t)\breve{\Sigma}_R^{-+}(\omega,t)\bigg]
\end{eqnarray}
Here the trace is over all the system orbitals and $\Sigma_{R}^{+-}(\omega,t)$,
etc. are (Wigner transformed) self-energy matrices due
to interaction with the right reservoir. $G^{+-}(\omega,t)$, etc. are Wigner
transforms of real time projections of system contour ordered Greens function
matrices with elements 
\begin{equation}
\label{eq-8}
 G^{c}_{mn}(\tau,\tau')=-\frac{i}{\hbar}\langle\mathcal{T}_c d_m(\tau)
d_n^\dag(\tau')\rangle
\end{equation}
where $\langle\cdots\rangle$ is the average with respect to the density matrix
evolving on Keldysh contour as defined in Eq.(\ref{eq-4}).
In Eq. (\ref{eq-7}) we have used the notation $\breve{A}^{-\eta}=-A^{-\eta}$ and
$A^{+\eta}=A^{+\eta}$, where $\eta=\pm$, to simplify the following expressions.
Expression for $\bigg[\frac{\partial\mbox{ln}{\cal
Z}(\lambda_L,\lambda_R,T-T_0)}{\partial(i\lambda_L)}\bigg]$ can be obtained by
replacing 'R' with 'L' in Eq.(\ref{eq-6}).
Hence all that we have to do to get the final expression for the GF is to
calculate the $\lambda(\tau)$-dependent Greens functions appearing in Eq.
(\ref{eq-7}). This is done in the following.\\
{\bf Approximation for the Greens function:}
The $\lambda$-dependent Greens function matrix defined on Keldysh contour (with
matrix elements in system orbital space defined in Eq.(\ref{eq-8})) satisfy the
following equation of motion \cite{Haug2008} (also known as left-Dyson equation)
on the Keldysh contour
\begin{equation}
\label{eq-10}
 \int_c
d\tau_1\big[G_0^{-1}(\tau,\tau_1)-\Sigma^{c}(\tau,\tau_1)\big]G^{c}(\tau_1,
\tau')=\delta^{c}(\tau,\tau')
\end{equation}
where
$\bigg[i\hbar\frac{\partial}{\partial\tau}-H_0(\tau)\bigg]G_0(\tau,
\tau')=\delta^{c}(\tau,\tau')$ and
$\Sigma^{c}(\tau,\tau')=\Sigma_{int}^{c}(\tau,\tau')+\Sigma_{leads}^{c}(\tau,
\tau')$
is the total self energy due to $H_{int}$ and system-reservoir coupling.
$\Sigma_{leads}^{c}(\tau,\tau')$ has matrix elements,
\begin{eqnarray}
\label{eq-11}
\Sigma^{c}_{leads,rr'}(\tau,\tau') &=& 
\sum_{\alpha}\displaystyle\sum_{k,k'} g_{\alpha
k,r}(\tau)e^{-i(\lambda_\alpha(\tau)-\lambda_\alpha(\tau'))}\nonumber\\
&\times&G^0_{\alpha k,\alpha k'}(\tau,\tau')g_{\alpha k',r'}^{*}(\tau').
\end{eqnarray}
Here $G^0_{\alpha k,\alpha k'}(\tau,\tau')$ is contour ordered Greens functions
of free reservoir. An explicit expression for $\Sigma_{int}^{c}$ depends on
$H_{int}$ which we keep general.
The above equation can be projected onto the real times to obtain equations of
motion for Keldysh matrix $\mathbb{\breve{G}}(t,t')$. By performing Wigner
transformation followed by Fourier transformation over quantum time, Eq.(10) can
be recast as \cite{Rammer2007}
 \begin{eqnarray}
\label{eq-12}
 &&\mathbb{\breve{G}}(\omega,t)=\mathbb{\breve{G}}_{ad}^{}(\omega,t)
 +\mathbb{\breve{G}}_{ad}^{}(\omega,t)\left[\mathbb{\breve{G}}_{ad}^{-1}(\omega,
t)\right.\nonumber\\
&&\left.\left(\mathbb{I}-\exp\big[-\frac{i}{2}[\overleftarrow{\partial}_t
\overrightarrow{\partial}_{\omega}-\overleftarrow{\partial}_{\omega}
\overrightarrow{\partial}_t ]\right)\mathbb{\breve{G}}(\omega,t)\right]
 \end{eqnarray}
where $\mathbb{I}$ is the identity matrix and $\mathbb{\breve{G}}(\omega,t)$ is
the system Greens function matrix Fourier transformed over quantum time.
$\overleftarrow{\partial}_t$, $\overrightarrow{\partial}_t$ represent classical
time $t$ derivative acting on the function to its left or to its right
respectively, similarly  $\overleftarrow{\partial}_{\omega}$,
$\overrightarrow{\partial}_{\omega}$ represent '$\omega$' derivatives.
The adiabatic contribution, $\mathbb{\breve{G}}_{ad}^{}$,  to the Greens
function satisfies the matrix equation,
 \begin{equation}
\label{eq-13}
\mathbb{\breve{G}}_{ad}^{}(\omega,t)=\bigg[\mathbb{\breve{G}}_{0}^{-1}(\omega,
t)-\breve{\Sigma}(\omega,t)\bigg]^{-1}.
 \end{equation}
This is similar to the usual steady state Greens function but the parameters are
replaced  with time dependent parameters. Here $\breve{\Sigma}(\omega,t)$ and
$\mathbb{\breve{G}}_0(\omega,t)$ are, respectively, the self energy and the non
interacting system Greens function matrices.
The second term in Eq.(\ref{eq-12}) represents a correction due to time
variation of the parameters. Note that Eq. (\ref{eq-12}) is exact. We solve
Eq.(\ref{eq-12}) to lowest order in time derivative by  iterating  the equation
for $\mathbb{\breve{G}}$ perturbatively in terms of time derivatives of
$\mathbb{\breve{G}}_{ad}^{}$ and retaining only terms linear in first derivative
in (classical) time.
 \begin{equation}
\label{eq-14}
 \mathbb{\breve{G}}(\omega,t)=\mathbb{\breve{G}}_{ad}^{}(\omega,t)+\frac{i}{2}
\mathbb{\breve{G}}_{ad}^{}(\omega,t)\{\mathbb{\breve{G}}_{ad}^{-1}(\omega,t),
\mathbb{\breve{G}}_{ad}(\omega,t)\}.
 \end{equation}
 where $\{,\}$ stand for Poisson bracket in $t$ and $\omega$ variables.
We note that both the left- and the right-Dyson equations lead to the same
approximate Eq. (\ref{eq-14}). We also note that Eq. (\ref{eq-14}) for
$\lambda=0$ case preserves all the symmetries (as a consequence of the unitary
evolution) of the Greens functions \cite{sup}.
Equation (\ref{eq-13}) and Eq. (\ref{eq-14}) can be solved to obtain
$\lambda$-dependent Greens functions which can be used to compute charge
transfer statistics using Eq.(\ref{eq-7}). In the following, we apply this
formalism to study Berry effect on the charge transfer statistics in a resonant
level model. \\
{\bf Adiabatically driven resonant level model:}
Consider a single electronic site connected via time-dependent hopping to two
electronic reservoirs. The Hamiltonian describing this model is the same as in
Eq. (\ref{eq-1}) with $H_{int}(t)=0$ and only single system orbital with time
independent energy $\epsilon=\omega_0$. We put $\hbar=1$ in this section. 
We compute the adiabatic Greens functions and the
lowest order non-adiabatic corrections using Eq.(\ref{eq-13}) and Eq.
(\ref{eq-14}) \cite{sup}. Using this Greens functions in Eq.(9),  we
obtain an expression for $\frac{1}{(T-T_0)}\bigg[\frac{\partial\mbox{ln}{\cal
Z}(\lambda_L,\lambda_R,T-T_0)}{\partial(i\lambda_R)}\bigg]$ \cite{sup}.
In order to emphasize the Berry phase effect, we consider the two reservoirs at
the same thermodynamic equilibrium  ($\mu_L=\mu_R=\mu$ i.e., zero external bias
and inverse temperatures, $\beta_L=\beta_R=\beta$. From $n_L$ and $n_R$ measurements, 
we can obtain the statistics of net charge, $n=(n_L-n_R)/2$,
transferred between the two reservoirs and the change in the charge, 
$N=n_L+n_R$, on the system. The transformation $(n_L,n_R)\leftrightarrow
(n,N)$ leads to $P(N,n,T-T_0)$ \cite{sup}.\\
{\it Statistics of particle change on the system:}
Since $\tilde{P}(N)=\sum_n\tilde{P}(N,n)$, the GF, 
$\tilde{Z}(\Lambda)$, for $\tilde{P}(N)$  satisfies 
$\frac{1}{(T-T_0)}\bigg[\frac{\partial\mbox{ln}{\cal\tilde{Z}}(\Lambda)}{
\partial(i\Lambda)}\bigg]=0$, thus $\tilde{Z}(\Lambda,T-T_0)=1$ and therefore,
$\tilde{P}(N,T-T_0)=\delta_{N0}$. This means that fluctuations in the electron
change on the system die in the long time limit. This is due to the finite
dimensionality of the system Hilbert space.\\
{\it Statistics of particles exchanged between reservoirs:}
Generating function $\tilde{Z}(\lambda)$ for the probability distribution
of $n$ electrons transferred from the right to the left reservoir, 
$\tilde{P}(n)=\sum_N\tilde{P}(N,n)$,  is obtained as \cite{sup}
 \begin{eqnarray}
\label{eq-15}
&&\frac{1}{(T-T_0)}\bigg[\frac{\partial\mbox{ln}{\cal
Z}(\lambda,T-T_0)}{\partial(i\lambda)}\bigg]=\frac{1}{T_p}\int_{0}^{T_p}dt\int_{
-\infty}^{+\infty}\frac{d\omega}{2\pi}\nonumber\\
&&\Bigg[\frac{\Gamma_L(t)\Gamma_R(t)f(\omega)(1-f(\omega))[e^{i\lambda}-e^{-i\lambda}]}{\Delta(\omega,t)}\nonumber\\
&+&\frac{[\dot{\Gamma}_L(t)+\dot{\Gamma}_R(t)]}{4[\Delta(\omega,t)]^2}(\omega-\omega_0)f'(\omega)
\bigg([\Gamma_R^2(t)-\Gamma_L^2(t)]\nonumber\\
&+&\Gamma_L(t)\Gamma_R(t)(1-2f(\omega))[e^{i\lambda}-e^{-i\lambda}]\bigg)\nonumber\\
&-&\frac{\Gamma_L(t)+\Gamma_R(t)}{2[\Delta(\omega,t)]^2}\bigg([\Gamma_L(t)\dot{\Gamma}_R(t)-\Gamma_R\dot{\Gamma}_L(t)]
f(\omega)(1-f(\omega))\nonumber\\
&&\big\{(0.5-f(\omega))+(\omega-\omega_0)f'(\omega)\big\}[e^{i\lambda}+e^{-i\lambda}-2]\bigg)\Bigg]
 \end{eqnarray}
where $f'(\omega)=\frac{\partial f(\omega)}{\partial \omega}$, $\Gamma_{\alpha}(t)=2\pi|g_{\alpha}(t)|^2\rho$,
$\dot{\Gamma}_{\alpha}(t)=\frac{\partial \Gamma_{\alpha}(t)}{\partial t}$,
$\Delta(\omega,t)=(\omega-\omega_0)^2+(\frac{\Gamma_L(t)+\Gamma_R(t)}{2}
)^2+\Gamma_L(t)\Gamma_R(t)f(\omega)(1-f(\omega))[e^{i\lambda}+e^{-i\lambda}-2]
$, and $T_p$ is the time period of driving which is assumed to be much larger
than the internal relaxation time of the system.
Here it is assumed that the second measurement is carried out 
after $q=\frac{T-T_0}{T_p}$
number of cycles  of driving (even if it is not the case
the error is insignificant at long measurement time, for which the present
formalism is developed).
The first integral in Eq. (\ref{eq-15}) is the so-called dynamical contribution
(similar to steady-state contribution with parameters replaced with time
dependent quantities with a time averaging). The second integral  can be
converted to parameter integral in $(\Gamma_L,\Gamma_R)$ space and is
independent of how fast the parameters are varied provided we are in the cyclic
adiabatic limit, thus it represents a Berry contribution.
When the two drivings are identical $\Gamma_L(t)=\Gamma_R(t)$, the Berry
contribution vanishes identically, the area traced in the parameter
space $(\Gamma_L,\Gamma_R)$ is zero.
Equation (\ref{eq-15}) allows us to compute the full statistics of net particles
transferred between the left and right reservoirs. For example, the expression
for the average pumped charge is obtained by setting $\lambda=0$ in Eq.
(\ref{eq-15}). The average number of electrons pumped per cycle is obtained as 
\begin{equation}
\label{eq-16}
N_{pump}=\frac{\beta}{8\pi^2}\int_0^{T_p}dt\partial_t(\Gamma_L(t)-\Gamma_R(t))\Im \Psi^{(1)}(Z)
\end{equation}
 where
$Z=\frac{1}{2}+i\frac{\beta}{2\pi}[\mu-\omega_0-i(\frac{\Gamma_L(t)+\Gamma_R(t)}{2})]$
and $\Im \Psi^{(1)}(Z)$ is the imaginary part of the trigamma function $\Psi^{(1)}(Z)$ 
of $Z$ \cite{Milne1972}.
It is clear that $N_{pump}=0$ when $\omega_0=\mu$ since $\Im \Psi^{(1)}(Z) =0$ in this case.
This result may be useful in identifying resonance
energy, $\omega_0$, of an unknown quantum system at the junction by applying an 
external gate voltage on the system such that the net pumped charge is zero. 
Indeed, also when $\Gamma_L(t)=\Gamma_R(t)$, there is no net pumping of electrons
between the two reservoirs. Thus the flux is purely driven due to Berry phase effects.
Additionally, we find that the flux changes sign from $\mu > \omega_0$ to $\mu <
\omega_0$. Similar behavior in the pumped charge was observed in Ref. \cite{Janine2007}.
This is demonstrated in the Fig. (\ref{fig-1}) for sinusoidal drivings.
Thus, both the phase difference between the drivings and the detuning act as 
driving forces that gives
rise to a net charge flux. Note that these forces are non-thermodynamic. The
Berry phase changes sign as the driving is reversed (the area in parameter space
is traced in reverse manner) and, as a consequence, the flux reverses the
direction. Note that the dynamical part in Eq. (\ref{eq-15}) only contributes
to the even cumulants. The Berry-phase part, on the other hand, in general, 
contributes to all the cumulants. However for even-cyclic drivings, $i.e.$,
$\Gamma_{L(R)}(-X)=\Gamma_{L(R)}(X)$ and  
$\Gamma_{L(R)}(t)\equiv\Gamma_{L(R)}(t/T_p\pm\phi/2)$, where $\phi$ is the
\begin{figure}[h]
\centering
\includegraphics[width=8cm]{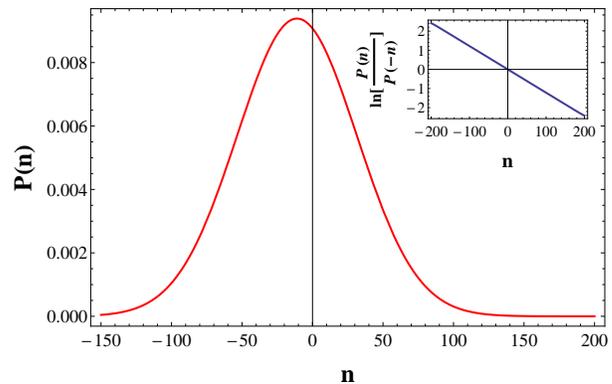}
\caption{Probability distribution for the net charge ($n$) transferred from the right to the left 
reservoir during the measurement time corresponding to $q=100$ with $T_p=100$, $\phi=\pi/2$, $\mu=0$,
and $\omega_0=3$. Time is in units of
$h\beta$. Drivings are the same as in Fig. (\ref{fig-1}).}
\label{fig-2}
\end{figure}
phase difference, the Berry part
contributes only to the odd cumulants. This fact may be helpful in designing
experiments to distinguish Berry contribution from the dynamical part. 
In Fig. (\ref{fig-2}) we present numerical result for the full distribution 
function $P(n)$ for measurement time corresponding to $q=100$. The distribution
seems symmetric (second and third cumulants are $\sim 18$ and $\sim 0.05$, respectively) 
and is peaked around the value $n\sim q N_{pump}$. In the inset
we show $\mbox{ln}(P(n)/P(-n))$ which grows linearly with $n$, confirming
the GC-type fluctuation symmetry. However this symmetry originates due to non-thermodynamic
forces and, as a consequence, the GFs for the forward (${\cal Z_F}$) and 
backward (${\cal Z_B}$) drivings satisfy, ${\cal Z_F}(\lambda)={\cal Z_B}(-\lambda)$) \cite{sup}.  
 
We developed an approximation scheme for computing the full counting statistics 
of adiabatic charge pumps within the NEGF formalism. We applied this formalism 
to study the statistics of pumped charge in a single level model where the 
coupling to the reservoirs are driven adiabatically in time.  
It is found that a net (non-zero) number of electrons can be transferred between 
two reservoirs kept at the
same thermodynamic states by adiabatically modulating the coupling strengths.  
An analytic expression is derived for the net charge pumped per
cycle entirely due to the geometric (Berry) phase. 
It is found that the phase difference between the drivings as
well as the energy difference between the level and reservoir chemical potential
(at zero bias) are the important parameters that determine the direction of the
pumped current. 
The statistics of the pumped charge is also influenced by the Berry-phase
and the corresponding distribution function follows the Gollavati-Cohen type 
symmetry. 

H. Y. and U. H. acknowledge the financial support from the Indian Institute of
Science, Bangalore, India.

\begin{widetext}

\section*{APPENDIX}

  \section{Derivation of Meir-Wingreen type expression for Generating function
[Eq.(6) of the main text]}

  Equation (5) of the main text can be recast in terms of the system Greens
function matrix alone by substituting the mixed Greens functions in terms of the
system Greens functions \cite{Haug2008}. We get,
\begin{eqnarray}
\label{App-eq-1}
\bigg[\frac{\partial\mbox{ln}{\cal
Z}(\lambda_L,\lambda_R,T-T_0)}{\partial(i\lambda_R)}\bigg]=\int_{T_0}^{T}
dt_1\int_{T_0}^{T}dt_2
\mbox{Tr}\bigg[\Sigma_{R}^{+-}(t_1,t_2)G_{}^{-+}(t_2,t_1)-G_{}^{+-}(t_1,
t_2)\Sigma_{R}^{-+}(t_2,t_1)\bigg]
\end{eqnarray}
Here the trace is over all the system orbitals and $\Sigma_{R}^{+-}(t,t')$, etc.
are real time projections of self-energy matrix $\Sigma_{R}(\tau,\tau')$ due to
interaction with the right reservoir. It has matrix elements, 
\begin{eqnarray}
\Sigma_{R;rr^\prime}(\tau,\tau^\prime)&=&\sum_{k,k'}
g_{Rk,r}(\tau)e^{-i(\lambda_R(\tau)-\lambda_R(\tau'))}
G^0_{Rk,Rk'}(\tau,\tau')g_{Rk',r'}^{*}(\tau')
\end{eqnarray}
where $G^0_{Rk,Rk'}(\tau,\tau')$ being contour ordered
Greens functions of free right reservoir).
$G^{+-}(t,t')$, etc. are real time projections of system contour ordered Greens
function matrix with elements defined in Eq.(7) of main text.
Now we use Wigner representation of the quantities in the integrand of Eq.(\ref{App-eq-1})
\cite{Rammer2007} i.e., we use the transformation pair
$A(t,t')=\int_{-\infty}^{+\infty} \frac{d\omega}{2\pi} \mathbb{A}(\omega,t_c)
e^{i\omega t_q}$ and $\mathbb{A}(\omega,t_c)=\int_{-\infty}^{+\infty} dt_q
A(t,t') e^{-i\omega t_q}$ (here $t_c=\frac{t+t'}{2}$ and $t_q=t-t'$ are
classical and quantum times respectively) to get
\begin{eqnarray}
\bigg[\frac{\partial\mbox{ln}{\cal Z}(\lambda_L,\lambda_R,T-T_0)}{\partial(i\lambda_R)}\bigg]
&=& \int_{T_0}^{T}dt_c \int_{-(T-T_0)}^{(T-T_0)}dt_q 
\int_{-\infty}^{+\infty}\frac{d\omega_1}{2\pi}\int_{-\infty}^{+\infty}\frac{d\omega_2}{2\pi}
e^{i(\omega_1-\omega_2)t_q} \nonumber\\
&&\mbox{Tr}\bigg[\Sigma_R^{+-}(\omega_1,t_c)\mathbb{G}^{-+}
(\omega_2,t_c)-\mathbb{G}^{+-}(\omega_1,t_c)\Sigma_R^{-+}(\omega_2,t_c)\bigg]
\end{eqnarray}
Here $t_c=\frac{t_1+t_2}{2}$ and $t_q=t_1-t_2$.We neglect the effect of
transients by assuming that the measurement time '$T-T_0$' is large compared to
internal relaxation times and driving time period (hence we send the $t_q$
integral from $-\infty$ to $\infty$). This leads to
\begin{eqnarray}
\label{App-eq-2}
\bigg[\frac{\partial\mbox{ln}{\cal Z}(\lambda_L,\lambda_R,T-T_0)}{\partial(i\lambda_R)}\bigg]=\int_{T_0}^{T}dt\int_
{-\infty}^{+\infty}\frac{d\omega}{2\pi}
\mbox{Tr}\bigg[\Sigma_R^{+-}(\omega,t)\mathbb{G}^{-+}(\omega,t)-\mathbb{G}^{+-
}(\omega,t)\Sigma_R^{-+}(\omega,t)\bigg].
\end{eqnarray}

\section{symmetrization of generating function}

From ${\cal Z}(\lambda_L,\lambda_R,T-T_0)$ we can obtain the combined 
distribution function $P(n_L,n_R,T-T_0)$ for the electron number change on left 
and right reservoirs over the measurement time period of $(T-T_0)$ using Eq.(2)
of the main text.
However in order to compute the statistics of net number ($N$) of electrons 
changed on both reservoirs (which is same as the number of electron changed on 
the system) and the net number (n) of electrons exchanged between both the 
reservoirs, we perform a coordinate transformation 
$(n_L,n_R)\rightarrow(N,n)=(\frac{n_L+n_R}{2},\frac{n_L-n_R}{2})$. Performing this 
transformation in Eq.(2) of main text, we obtain
\begin{eqnarray}
  \tilde{P}(N,n,T-T_0)=\frac{1}{2}\int_0^{2\pi}\frac{d\Lambda}{2\pi}\int_{-2\pi}^{2\pi}\frac{d\lambda}{2\pi}
{\cal\tilde{Z}}(\Lambda,\lambda,T-T_0) e^{i(\Lambda N + \lambda n)},
\end{eqnarray}
 where the factor $1/2$ appears due to the Jacobian of transformation.
 ${\cal \tilde{Z}}(\Lambda,\lambda,T-T_0)$ is obtained from ${\cal
Z}(\lambda_L,\lambda_R,T-T_0)$ by performing  a coordinate transformation 
$(\lambda_L,\lambda_R)\rightarrow(\Lambda,\lambda)=(\lambda_L+\lambda_R,\lambda_L-\lambda_R)$.
From $\tilde{P}(N,n,T-T_0)$ we can get $\tilde{P}(N,T-T_0)$ or 
$\tilde{P}(n,T-T_0)$ by summing over 'n' or 'N' respectively.

Summing over $N$, we get
\begin{eqnarray}
  \tilde{P}(n,T-T_0)&=&\frac{1}{2}\int_{-2\pi}^{2\pi}\frac{d\lambda}{2\pi}
{\cal\tilde{Z}}(\lambda,T-T_0) e^{i\lambda n}
= \int_{0}^{2\pi}\frac{d\lambda}{2\pi}
{\cal\tilde{Z}}(\lambda,T-T_0) e^{i\lambda n},
\end{eqnarray}
where the second equality follows from the $2\pi$ periodicity of ${\cal\tilde{Z}}(\lambda,T-T_0)$.

\section{ $\lambda$-dependent Wigner transformed Greens functions for the resonant level model}

In this section we describe the procedure to get $\lambda$-dependent Greens
functions (up to lowest linear order correction in driving) for the resonant
level model.
We calculate Greens functions on the Keldysh contour which goes from $-\infty$
to $\infty$ and back to $\infty$ under large measurement time assumption in
order to neglect initial correlations.

 The non-interacting Greens function $\mathbb{\breve{G}}_0$ is defined below Eq. (8) in the
 main text. The inverse of $\mathbb{\breve{G}}_0$ in Wigner transformed system is obtained as
 \begin{align}
  \mathbb{\breve{G}}_0^{-1}(\omega,t)=
  \begin{pmatrix}
  \omega-\omega_0+i\eta & 0\\
   0 & \omega-\omega_0\nonumber-i\eta\\
 \end{pmatrix}
 \end{align}
 where $\eta=0^+$. The $\lambda$-dependent Wigner transformed self-energy (with terms up to linear in
first derivative in classical time) due to coupling to the reservoirs is 
 $\breve{\Sigma}(\omega,t)=\sum_{\alpha=L,R} \breve{\Sigma}_{\alpha}(\omega,t)$ with
  \begin{align}
           \breve{\Sigma}_{\alpha}(\omega,t)=
  \begin{pmatrix}
  -i\Gamma_{\alpha}(t)(0.5-f_{\alpha}(\omega)) & i\Gamma_{\alpha}(t)
f_{\alpha}(\omega) e^{i\lambda_{\alpha}} \\
   i\Gamma_{\alpha}(t) (1-f_{\alpha}(\omega)) e^{-i\lambda_{\alpha}} &
i\Gamma_{\alpha}(t)(0.5-f_{\alpha}(\omega)) \nonumber\\
  \end{pmatrix}
 \end{align}
 Here we assume that the system-reservoir couplings are real and independent of energy (wide-band approximation). 
 The first order correction to the self-energy due to external driving is then zero.

 Using Eq.(11) of the main text together with the above two equations for 
 $\mathbb{\breve{G}}_0^{-1}$ and $\breve{\Sigma}_{\alpha}$,
 we get an expression for the adiabatic Greens function as 

   \begin{align}
\mathbb{\breve{G}}_{ad}(\omega,t)=\frac{1}{\Delta(\omega,t)}
\begin{pmatrix}
\omega-\omega_0-i\Gamma_L(t)(0.5-f_L(\omega))-i\Gamma_R(t)(0.5-f_R(\omega)) &
i\Gamma_L(t) f_L(\omega) e^{i\lambda_L}+i\Gamma_R(t) f_R(\omega)
e^{i\lambda_R}\\
i\Gamma_L(t) (1-f_L(\omega)) e^{-i\lambda_L}+i\Gamma_R(t) (1-f_R(\omega))
e^{-i\lambda_R} &
\omega-\omega_0+i\Gamma_L(t)(0.5-f_L(\omega))+i\Gamma_R(t)(0.5-f_R(\omega)) 
\nonumber\\
\end{pmatrix}
 \end{align}

  with
\begin{eqnarray}
  \Delta(\omega,t)=(\omega-\epsilon)^2+(\frac{\Gamma_L(t)+\Gamma_R(t)}{2}
)^2+\Gamma_L(t)\Gamma_R(t)[f_L(\omega)(1-f_R(\omega))(e^{i(\lambda_L-\lambda_R)}
-1)+f_R(\omega)(1-f_L(\omega))(e^{-i(\lambda_L-\lambda_R)}-1)]\nonumber
\end{eqnarray}
where $f_L(\omega)$ and $f_R(\omega)$ are Fermi functions of the left and right
reservoirs, respectively.
 
 Using $\mathbb{\breve{G}}_{ad}(\omega,t)$ in Eq.(12) of main text we calculate
the lowest order correction to the Greens functions. We give expressions only
for '$+-$' and '$-+$' components.
 
 \begin{eqnarray}
 \mathbb{\breve{G}}^{+-}(\omega,t)&=&\frac{i\Gamma_L(t) f_L(\omega)
e^{i\lambda_L}+i\Gamma_R(t) f_R(\omega)
e^{i\lambda_R}}{\Delta(\omega,t)}-\frac{i[\dot{\Gamma}_L+\dot{\Gamma}_R]
(\omega-\omega_0)}{2[\Delta(\omega,t)]^2}\big[\Gamma_Lf'_L(\omega)e^{i\lambda_L}
+\Gamma_Rf'_R(\omega)e^{i\lambda_R}\big]\nonumber\\
 &&+\frac{i(\Gamma_L(t)\dot{\Gamma}_R(t)-\Gamma_R(t)\dot{\Gamma}_L(t))}{2[
\Delta(\omega,t)]^2}\big[(1-2f_R(\omega))f_L(\omega)e^{i\lambda_L}
-(1-2f_L(\omega))f_R(\omega)e^{i\lambda_R}\big]\nonumber\\
 &&+\frac{i(\omega-\omega_0)}{[\Delta(\omega,t)]^2}\big[\Gamma_L(t)\dot{\Gamma}
_R(t)f_R(\omega)f'_L(\omega)-\Gamma_R(t)\dot{\Gamma}
_L(t)f_L(\omega)f'_R(\omega)\big](e^{i\lambda_L}-e^{i\lambda_R})\\
  \mathbb{\breve{G}}^{-+}(\omega,t)&=&\frac{i\Gamma_L(t) (1-f_L(\omega))
e^{-i\lambda_L}+i\Gamma_R(t) (1-f_R(\omega))
e^{-i\lambda_R}}{\Delta(\omega,t)}+\frac{i[\dot{\Gamma}_L+\dot{\Gamma}_R]
(\omega-\omega_0)}{2[\Delta(\omega,t)]^2}\big[\Gamma_Lf'_L(\omega)e^{-i\lambda_L
}+\Gamma_Rf'_R(\omega)e^{-i\lambda_R}\big]\nonumber\\
 &&-\frac{i(\Gamma_L(t)\dot{\Gamma}_R(t)-\Gamma_R(t)\dot{\Gamma}_L(t))}{2[
\Delta(\omega,t)]^2}\big[(1-2f_R(\omega))(1-f_L(\omega))e^{-i\lambda_L}
-(1-2f_L(\omega))(1-f_R(\omega))e^{-i\lambda_R}\big]\nonumber\\
 &&-\frac{i(\omega-\omega_0)}{[\Delta(\omega,t)]^2}\big[\Gamma_L(t)\dot{\Gamma}
_R(t)(1-f_R(\omega))f'_L(\omega)-\Gamma_R(t)\dot{\Gamma}
_L(t)(1-f_L(\omega))f'_R(\omega)\big](e^{-i\lambda_L}-e^{-i\lambda_R})
 \end{eqnarray}

Using $\mathbb{\breve{G}}^{+-}(\omega,t)$, $\mathbb{\breve{G}}^{-+}(\omega,t)$,
$\breve{\Sigma}^{+-}_R(\omega,t)$ and $\breve{\Sigma}^{-+}_R(\omega,t)$ in
Eq.(\ref{App-eq-2}) we get an expression for
$\frac{1}{(T-T_0)}\bigg[\frac{\partial\mbox{ln}{\cal
Z}(\lambda_L,\lambda_R,T-T_0)}{\partial(i\lambda_R)}\bigg]$ to lowest order
correction,
\begin{eqnarray}
\label{App-eq-3}
 &&\frac{1}{(T-T_0)}\bigg[\frac{\partial\mbox{ln}{\cal
Z}(\lambda_L,\lambda_R,T-T_0)}{\partial(i\lambda_R)}\bigg]=\nonumber\\ 
 &&\frac{1}{T_p}\int_{0}^{T_p}dt\int_{-\infty}^{+\infty}\frac{d\omega}{2\pi}
\frac{\Gamma_L(t)\Gamma_R(t)}{\Delta(\omega,t)}[f_R(\omega)(1-f_L(\omega))e^{
-i(\lambda_L-\lambda_R)}-f_L(\omega)(1-f_R(\omega))e^{i(\lambda_L-\lambda_R)}]
\nonumber\\
 &&+\frac{1}{T_p}\int_{0}^{T_p}dt\int_{-\infty}^{+\infty}\frac{d\omega}{2\pi}
\frac{1}{[\Delta(\omega,t)]^2}\times\nonumber\\
 &&\bigg[\frac{\Gamma_R(t)[\Gamma_L(t)\dot{\Gamma}_R(t)-\Gamma_R(t)\dot{\Gamma}
_L(t)]}{2}[f_L(\omega)-f_R(\omega)]-\frac{\Gamma_R(t)[\dot{\Gamma}_L(t)+\dot{
\Gamma}_R(t)](\omega-\omega_0)}{2}[
\Gamma_L(t)f'_L(\omega)+\Gamma_R(t)f'_R(\omega)]\nonumber\\
 &&+\frac{\Gamma_R(t)[\Gamma_L(t)\dot{\Gamma}_R(t)-\Gamma_R(t)\dot{\Gamma}_L(t)]
}{2}(1-2f_R(\omega))[f_L(\omega)(1-f_R(\omega))(e^{i(\lambda_L-\lambda_R)}
-1)+f_R(\omega)(1-f_L(\omega))(e^{-i(\lambda_L-\lambda_R)}-1)]\nonumber\\
 &&-\frac{\Gamma_L(t)\Gamma_R(t)[\dot{\Gamma}_L(t)+\dot{\Gamma}_R(t)]
(\omega-\omega_0)}{2}f'_L(\omega)[(1-f_R(\omega))(e^{i(\lambda_L-\lambda_R)}
-1)+f_R(\omega)(e^{-i(\lambda_L-\lambda_R)}-1)]\nonumber\\
 &&+\Gamma_L(t)\Gamma_R(t)\dot{\Gamma}
_R(t)(\omega-\epsilon)f'_L(\omega)f_R(\omega)(1-f_R(\omega))[e^{
i(\lambda_L-\lambda_R)}+e^{-i(\lambda_L-\lambda_R)}-2]\nonumber\\
 &&-\Gamma_R^2(t)\dot{\Gamma}_L(t)f'_R(\omega)(\omega-\omega_0)[
f_L(\omega)(1-f_R(\omega))(e^{i(\lambda_L-\lambda_R)}
-1)+f_R(\omega)(1-f_L(\omega))(e^{-i(\lambda_L-\lambda_R)}-1)]\bigg]
 \end{eqnarray}
where $f'_{\alpha}(\omega)$ is $\omega$ derivative of $f'_{\alpha}(\omega)$,
$\dot{\Gamma}_{\alpha}(t)$ is time derivative of $\Gamma_{\alpha}(t)$ and
'$T_p$' is
the time period of driving which is assumed to be much larger than the internal
relaxation time of the system. A similar expression for 
$\frac{1}{(T-T_0)}\bigg[\frac{\partial\mbox{ln}{\cal
Z}(\lambda_L,\lambda_R,T-T_0)}{\partial(i\lambda_L)}\bigg]$ can be obtained by
swapping 'L' and 'R' labels in 
Eq. (\ref{App-eq-3}). Here it is assumed that the second measurement is carried out after
an integer number of cycles $n=\frac{T-T_0}{T_p}$ of driving has been performed 
(even if it is not the case the error is minimal for large time statistics, for
which the present formalism is developed). Using expressions for
$\frac{1}{(T-T_0)}\bigg[\frac{\partial\mbox{ln}{\cal
Z}(\lambda_L,\lambda_R,T-T_0)}{\partial(i\lambda_R)}\bigg]$ and
$\frac{1}{(T-T_0)}\bigg[\frac{\partial\mbox{ln}{\cal
Z}(\lambda_L,\lambda_R,T-T_0)}{\partial(i\lambda_L)}\bigg]$, 
expression for the generating function for particle change on system and
particle exchanged between reservoirs can be obtained. These expressions are
presented in the main text for $f_L(\omega)=f_R(\omega)=f(\omega)$ case.

\section{Symmetries of the Greens functions}

As the full system is evolving with respect to a Hamiltonian, the evolution is
unitary when $\lambda_L=\lambda_R=0$. As a consequence, matrix elements of 
$\bf{\mathbb{\breve{G}}(\tau,\tau')}$ satisfy the symmetries \cite{Rammer2007}:
(i)
$[\mathbb{\breve{G}}_{mn}^{++}(t,t')]^{*}=\mathbb{\breve{G}}_{nm}^{--}(t',t)$,
(ii)
$[\mathbb{\breve{G}}_{mn}^{--}(t,t')]^{*}=\mathbb{\breve{G}}_{nm}^{++}(t',t)$,
(iii)
$[\mathbb{\breve{G}}_{mn}^{+-}(t,t')]^{*}=-\mathbb{\breve{G}}_{nm}^{+-}(t',t)$
and
(iv)
$[\mathbb{\breve{G}}_{mn}^{-+}(t,t')]^{*}=-\mathbb{\breve{G}}_{nm}^{-+}(t',t)$.
These symmetries in Wigner representation can be summarized in a matrix form as
\begin{eqnarray}
 &&\bf{\mathbb{\breve{G}}}(\omega,t)=
 \begin{bmatrix}
    \mathbb{\breve{G}}^{++}(\omega,t)&\mathbb{\breve{G}}^{+-}(\omega,t)\\
  \mathbb{\breve{G}}^{-+}(\omega,t)&\mathbb{\breve{G}}^{--}(\omega,t)\nonumber\\
   \end{bmatrix} \Longrightarrow
  [\bf{\mathbb{\breve{G}}}(\omega,t)]^{*}=
  \begin{bmatrix}
  [\mathbb{\breve{G}}^{--}(\omega,t)]^T&-[\mathbb{\breve{G}}^{+-}(\omega,t)]^T\\
-[\mathbb{\breve{G}}^{-+}(\omega,t)]^T&[\mathbb{\breve{G}}^{++}(\omega,t)]
^T\nonumber\\
  \end{bmatrix}
 \end{eqnarray}
The approximate Greens function matrix used in this work can be expressed as
 \begin{eqnarray} 
\mathbb{\breve{G}}(\omega,t)=\mathbb{\breve{G}}_{ad}^{}(\omega,t)+\frac{i}{2}
\bigg[\big\{\partial_{\omega}\mathbb{\breve{G}}_{ad}^{}(\omega,t)\big\}\mathbb{\breve{G}}_
{ad}^{-1}(\omega,t)\partial_{t}\mathbb{\breve{G}}_{ad}(\omega,t)
-\big\{\partial_{t}\mathbb{\breve{G}}_{ad}^{}(\omega,t)\big\}\mathbb{\breve{G}}_{ad}^{-1}
(\omega,t)\partial_{\omega}\mathbb{\breve{G}}_{ad}(\omega,t)\bigg]
 \end{eqnarray}
This approximate Greens function satisfies all the above symmetries provided
$\mathbb{\breve{G}}_{ad}^{}(\omega,t)$ satisfies the above symmetries which
in turn 
requires $\mathbb{\breve{G}}_{0}(\omega,t)$ and $\bf{\breve{\Sigma}}(\omega,t)$
to satisfy the above symmetries, which they do. Another important symmetry is 
$\mathbb{\breve{G}}_{mn}^{++}(t,t')-\mathbb{\breve{G}}_{mn}^{--}(t,t')=\mathbb{
\breve{G}}_{mn}^{+-}(t,t')-\mathbb{\breve{G}}_{mn}^{-+}(t,t')$ which in 
Wigner representation becomes
$\mathbb{\breve{G}}_{mn}^{++}(\omega,t)-\mathbb{\breve{G}}_{mn}^{--}(\omega,
t)=\mathbb{\breve{G}}_{mn}^{+-}(\omega,t)-\mathbb{\breve{G}}_{mn}^{-+}(t\omega,
t)$ 
is also satisfied by the above approximate Greens function provided
$\mathbb{\breve{G}}_{0}(\omega,t)$ and $\breve{\Sigma}(\omega,t)$ also satisfy
it, which they do.

\section{Detailed fluctuation theorem for pumped charge}

We consider a special case when both the reservoirs are at the same thermodynamic states (i.e., 
$\beta_L=\beta_R=\beta$ and $\mu_L=\mu_R=\mu$) and drivings are of the form 
$\Gamma_L(t)=\Gamma(\frac{2\pi t}{T_p}-\frac{\phi}{2})$ and $\Gamma_R(t)=\Gamma(\frac{2\pi t}{T_p}+\frac{\phi}{2})$
(where $\Gamma(x)$ is an even periodic function with period $2\pi$). During the time-reversed
evolution, the drivings at time $t$ are obtained by substituting $t\rightarrow T_p-t$ for drivings in the forward evolution.
Clearly the Hamiltonian does not have time-reversal symmetry as $\Gamma_{L(R)}(T_p-t)=\Gamma_{R(L)}(t)$, the left and right couplings switch roles. 
Then using Eq. (13) of the main text, 
it can be shown that, ${\cal \tilde{Z}_F}(\lambda,T-T_0)={\cal \tilde{Z}_B}(-\lambda,T-T_0)$, where ${\cal \tilde{Z}_F}$ and ${\cal \tilde{Z}_B}$ are moment generating functions of probability distribution function for 
number of particles exchanged between two reservoirs with forward and backward driving protocols, respectively.
This is the Gallavotti-Cohen type symmetry for zero bias case. This symmetry leads to a detailed fluctuation theorem (FT) 
\begin{equation}
\label{App-eq-4}
\lim_{(T-T_0)\rightarrow \infty} \frac{P_F(n,T-T_0)}{P_B(-n,T-T_0)} = 1,
\end{equation}
which is consistent with the standard (non-driven) steady-state FT for charge transfer in single resonant level system
at equilibrium (zero external bias). Result in Eq. (\ref{App-eq-4}) is a consequence of the 
${\cal P}\otimes\Theta$, where ${\cal P}$ and $\Theta$ are the parity (left $\leftrightarrow$ right) 
and the time-reversal operations,
symmetry of the Hamiltonian at zero bias and the drivings considered above.
However, at steady-state $P_F \equiv P_B$, and the above relation leads
to $P_F(n) = P_F(-n)$ at large measurement times. However, for a driven case, as we show in the main text,
we  find that  ${\cal \tilde{Z}_F}(\lambda,T-T_0)\neq {\cal \tilde{Z}_F}(-\lambda,T-T_0)$
due to Berry phase which leads to a finite net charge transfer between reservoirs .

\end{widetext}

\bibliography{citation.bib}

\begin{thebibliography}{28}
\expandafter\ifx\csname natexlab\endcsname\relax\def\natexlab#1{#1}\fi
\expandafter\ifx\csname bibnamefont\endcsname\relax
  \def\bibnamefont#1{#1}\fi
\expandafter\ifx\csname bibfnamefont\endcsname\relax
  \def\bibfnamefont#1{#1}\fi
\expandafter\ifx\csname citenamefont\endcsname\relax
  \def\citenamefont#1{#1}\fi
\expandafter\ifx\csname url\endcsname\relax
  \def\url#1{\texttt{#1}}\fi
\expandafter\ifx\csname urlprefix\endcsname\relax\def\urlprefix{URL }\fi
\providecommand{\bibinfo}[2]{#2}
\providecommand{\eprint}[2][]{\url{#2}}

\bibitem[{\citenamefont{{Berry}}(1984)}]{Berry1984}
\bibinfo{author}{\bibfnamefont{M.~V.} \bibnamefont{{Berry}}},
  \bibinfo{journal}{Royal Society of London Proceedings Series A}
  \textbf{\bibinfo{volume}{392}}, \bibinfo{pages}{45} (\bibinfo{year}{1984}).

\bibitem[{\citenamefont{Chruscinski and Jamiolkowski}(2012)}]{Chruscinski2012}
\bibinfo{author}{\bibfnamefont{D.}~\bibnamefont{Chruscinski}} \bibnamefont{and}
  \bibinfo{author}{\bibfnamefont{A.}~\bibnamefont{Jamiolkowski}},
  \emph{\bibinfo{title}{Geometric phases in classical and quantum mechanics}},
  vol.~\bibinfo{volume}{36} (\bibinfo{publisher}{Springer Science \& Business
  Media}, \bibinfo{year}{2012}).

\bibitem[{\citenamefont{Bohm et~al.}(2013)\citenamefont{Bohm, Mostafazadeh,
  Koizumi, Niu, and Zwanziger}}]{Bohm2013}
\bibinfo{author}{\bibfnamefont{A.}~\bibnamefont{Bohm}},
  \bibinfo{author}{\bibfnamefont{A.}~\bibnamefont{Mostafazadeh}},
  \bibinfo{author}{\bibfnamefont{H.}~\bibnamefont{Koizumi}},
  \bibinfo{author}{\bibfnamefont{Q.}~\bibnamefont{Niu}}, \bibnamefont{and}
  \bibinfo{author}{\bibfnamefont{J.}~\bibnamefont{Zwanziger}},
  \emph{\bibinfo{title}{The Geometric Phase in Quantum Systems: Foundations,
  Mathematical Concepts, and Applications in Molecular and Condensed Matter
  Physics}} (\bibinfo{publisher}{Springer Science \& Business Media},
  \bibinfo{year}{2013}).

\bibitem[{\citenamefont{Cheung et~al.}(1988)\citenamefont{Cheung, Gefen,
  Riedel, and Shih}}]{Cheung1988}
\bibinfo{author}{\bibfnamefont{H.-F.} \bibnamefont{Cheung}},
  \bibinfo{author}{\bibfnamefont{Y.}~\bibnamefont{Gefen}},
  \bibinfo{author}{\bibfnamefont{E.~K.} \bibnamefont{Riedel}},
  \bibnamefont{and} \bibinfo{author}{\bibfnamefont{W.-H.} \bibnamefont{Shih}},
  \bibinfo{journal}{Phys. Rev. B} \textbf{\bibinfo{volume}{37}},
  \bibinfo{pages}{6050} (\bibinfo{year}{1988}).

\bibitem[{\citenamefont{Loss et~al.}(1990)\citenamefont{Loss, Goldbart, and
  Balatsky}}]{Balatski1990}
\bibinfo{author}{\bibfnamefont{D.}~\bibnamefont{Loss}},
  \bibinfo{author}{\bibfnamefont{P.}~\bibnamefont{Goldbart}}, \bibnamefont{and}
  \bibinfo{author}{\bibfnamefont{A.~V.} \bibnamefont{Balatsky}},
  \bibinfo{journal}{Phys. Rev. Lett.} \textbf{\bibinfo{volume}{65}},
  \bibinfo{pages}{1655} (\bibinfo{year}{1990}).

\bibitem[{\citenamefont{Splettstoesser
  et~al.}(2003)\citenamefont{Splettstoesser, Governale, and
  Z{\"u}licke}}]{Splettstoesser2003}
\bibinfo{author}{\bibfnamefont{J.}~\bibnamefont{Splettstoesser}},
  \bibinfo{author}{\bibfnamefont{M.}~\bibnamefont{Governale}},
  \bibnamefont{and}
  \bibinfo{author}{\bibfnamefont{U.}~\bibnamefont{Z{\"u}licke}},
  \bibinfo{journal}{Phys. Rev. B} \textbf{\bibinfo{volume}{68}},
  \bibinfo{pages}{165341} (\bibinfo{year}{2003}).

\bibitem[{\citenamefont{Zhou et~al.}(2003)\citenamefont{Zhou, Cho, and
  McKenzie}}]{Zhou2003}
\bibinfo{author}{\bibfnamefont{H.-Q.} \bibnamefont{Zhou}},
  \bibinfo{author}{\bibfnamefont{S.~Y.} \bibnamefont{Cho}}, \bibnamefont{and}
  \bibinfo{author}{\bibfnamefont{R.~H.} \bibnamefont{McKenzie}},
  \bibinfo{journal}{Phys. Rev. Lett.} \textbf{\bibinfo{volume}{91}},
  \bibinfo{pages}{186803} (\bibinfo{year}{2003}).

\bibitem[{\citenamefont{Albash et~al.}(2012)\citenamefont{Albash, Boixo, Lidar,
  and Zanardi}}]{Albash2012}
\bibinfo{author}{\bibfnamefont{T.}~\bibnamefont{Albash}},
  \bibinfo{author}{\bibfnamefont{S.}~\bibnamefont{Boixo}},
  \bibinfo{author}{\bibfnamefont{D.~A.} \bibnamefont{Lidar}}, \bibnamefont{and}
  \bibinfo{author}{\bibfnamefont{P.}~\bibnamefont{Zanardi}},
  \bibinfo{journal}{New J. Phys.} \textbf{\bibinfo{volume}{14}},
  \bibinfo{pages}{123016} (\bibinfo{year}{2012}).

\bibitem[{\citenamefont{Harbola et~al.}(2006)\citenamefont{Harbola, Esposito,
  and Mukamel}}]{Harbola2006}
\bibinfo{author}{\bibfnamefont{U.}~\bibnamefont{Harbola}},
  \bibinfo{author}{\bibfnamefont{M.}~\bibnamefont{Esposito}}, \bibnamefont{and}
  \bibinfo{author}{\bibfnamefont{S.}~\bibnamefont{Mukamel}},
  \bibinfo{journal}{Phys. Rev. B} \textbf{\bibinfo{volume}{74}},
  \bibinfo{pages}{235309} (\bibinfo{year}{2006}).

\bibitem[{\citenamefont{Ren et~al.}(2010)\citenamefont{Ren, H{\"a}nggi, Li
  et~al.}}]{Ren2010}
\bibinfo{author}{\bibfnamefont{J.}~\bibnamefont{Ren}},
  \bibinfo{author}{\bibfnamefont{P.}~\bibnamefont{H{\"a}nggi}},
  \bibinfo{author}{\bibfnamefont{B.}~\bibnamefont{Li}}, \bibnamefont{et~al.},
  \bibinfo{journal}{Phys. Rev. Lett.} \textbf{\bibinfo{volume}{104}},
  \bibinfo{pages}{170601} (\bibinfo{year}{2010}).

\bibitem[{\citenamefont{Segal}(2008)}]{Segal2008}
\bibinfo{author}{\bibfnamefont{D.}~\bibnamefont{Segal}},
  \bibinfo{journal}{Phys. Rev. Lett.} \textbf{\bibinfo{volume}{101}},
  \bibinfo{pages}{260601} (\bibinfo{year}{2008}).

\bibitem[{\citenamefont{Yuge et~al.}(2012)\citenamefont{Yuge, Sagawa, Sugita,
  and Hayakawa}}]{Yuge2012}
\bibinfo{author}{\bibfnamefont{T.}~\bibnamefont{Yuge}},
  \bibinfo{author}{\bibfnamefont{T.}~\bibnamefont{Sagawa}},
  \bibinfo{author}{\bibfnamefont{A.}~\bibnamefont{Sugita}}, \bibnamefont{and}
  \bibinfo{author}{\bibfnamefont{H.}~\bibnamefont{Hayakawa}},
  \bibinfo{journal}{Phys. Rev. B} \textbf{\bibinfo{volume}{86}},
  \bibinfo{pages}{235308} (\bibinfo{year}{2012}).

\bibitem[{\citenamefont{Sinitsyn}(2009)}]{Sinitsyn2009}
\bibinfo{author}{\bibfnamefont{N.}~\bibnamefont{Sinitsyn}},
  \bibinfo{journal}{J. Phys. A: Math. Theo.} \textbf{\bibinfo{volume}{42}},
  \bibinfo{pages}{193001} (\bibinfo{year}{2009}).

\bibitem[{\citenamefont{Brouwer}(1998)}]{Brouwer1998}
\bibinfo{author}{\bibfnamefont{P.}~\bibnamefont{Brouwer}},
  \bibinfo{journal}{Phys. Rev. B} \textbf{\bibinfo{volume}{58}},
  \bibinfo{pages}{R10135} (\bibinfo{year}{1998}).

\bibitem[{\citenamefont{Switkes et~al.}(1999)\citenamefont{Switkes, Marcus,
  Campman, and Gossard}}]{Switkes1999}
\bibinfo{author}{\bibfnamefont{M.}~\bibnamefont{Switkes}},
  \bibinfo{author}{\bibfnamefont{C.}~\bibnamefont{Marcus}},
  \bibinfo{author}{\bibfnamefont{K.}~\bibnamefont{Campman}}, \bibnamefont{and}
  \bibinfo{author}{\bibfnamefont{A.}~\bibnamefont{Gossard}},
  \bibinfo{journal}{Science} \textbf{\bibinfo{volume}{283}},
  \bibinfo{pages}{1905} (\bibinfo{year}{1999}).

\bibitem[{\citenamefont{Kita}(2010)}]{Kita2010}
\bibinfo{author}{\bibfnamefont{T.}~\bibnamefont{Kita}}, \bibinfo{journal}{Prog.
  Theo. Phys.} \textbf{\bibinfo{volume}{123}}, \bibinfo{pages}{581}
  (\bibinfo{year}{2010}).

\bibitem[{\citenamefont{Esposito et~al.}(2009)\citenamefont{Esposito, Harbola,
  and Mukamel}}]{Esposito2009}
\bibinfo{author}{\bibfnamefont{M.}~\bibnamefont{Esposito}},
  \bibinfo{author}{\bibfnamefont{U.}~\bibnamefont{Harbola}}, \bibnamefont{and}
  \bibinfo{author}{\bibfnamefont{S.}~\bibnamefont{Mukamel}},
  \bibinfo{journal}{Rev. Mod. Phys.} \textbf{\bibinfo{volume}{81}},
  \bibinfo{pages}{1665} (\bibinfo{year}{2009}).

\bibitem[{\citenamefont{Andreev and Kamenev}(2000)}]{Andreev2000}
\bibinfo{author}{\bibfnamefont{A.}~\bibnamefont{Andreev}} \bibnamefont{and}
  \bibinfo{author}{\bibfnamefont{A.}~\bibnamefont{Kamenev}},
  \bibinfo{journal}{Phys. Rev. Lett.} \textbf{\bibinfo{volume}{85}},
  \bibinfo{pages}{1294} (\bibinfo{year}{2000}).

\bibitem[{\citenamefont{Levitov}(2001)}]{Levitov2001}
\bibinfo{author}{\bibfnamefont{L.}~\bibnamefont{Levitov}},
  \bibinfo{journal}{arXiv preprint cond-mat/0103617}  (\bibinfo{year}{2001}).

\bibitem[{\citenamefont{Moskalets and B{\"u}ttiker}(2004)}]{Moskalets2004}
\bibinfo{author}{\bibfnamefont{M.}~\bibnamefont{Moskalets}} \bibnamefont{and}
  \bibinfo{author}{\bibfnamefont{M.}~\bibnamefont{B{\"u}ttiker}},
  \bibinfo{journal}{Phys. Rev. B} \textbf{\bibinfo{volume}{70}},
  \bibinfo{pages}{245305} (\bibinfo{year}{2004}).

\bibitem[{\citenamefont{Rammer}(2007)}]{Rammer2007}
\bibinfo{author}{\bibfnamefont{J.}~\bibnamefont{Rammer}},
  \emph{\bibinfo{title}{Quantum field theory of non-equilibrium states}}
  (\bibinfo{publisher}{Cambridge University Press}, \bibinfo{year}{2007}).

\bibitem[{\citenamefont{Haug et~al.}(2008)\citenamefont{Haug, Jauho, and
  Cardona}}]{Haug2008}
\bibinfo{author}{\bibfnamefont{H.}~\bibnamefont{Haug}},
  \bibinfo{author}{\bibfnamefont{A.-P.} \bibnamefont{Jauho}}, \bibnamefont{and}
  \bibinfo{author}{\bibfnamefont{M.}~\bibnamefont{Cardona}},
  \emph{\bibinfo{title}{Quantum kinetics in transport and optics of
  semiconductors}}, vol.~\bibinfo{volume}{2} (\bibinfo{publisher}{Springer},
  \bibinfo{year}{2008}).

\bibitem[{\citenamefont{Gogolin and Komnik}(2006)}]{Gogolin2006}
\bibinfo{author}{\bibfnamefont{A.}~\bibnamefont{Gogolin}} \bibnamefont{and}
  \bibinfo{author}{\bibfnamefont{A.}~\bibnamefont{Komnik}},
  \bibinfo{journal}{Phys. Rev. B} \textbf{\bibinfo{volume}{73}},
  \bibinfo{pages}{195301} (\bibinfo{year}{2006}).

\bibitem[{\citenamefont{Bagrets and Nazarov}(2003)}]{Bagrets2003}
\bibinfo{author}{\bibfnamefont{D.}~\bibnamefont{Bagrets}} \bibnamefont{and}
  \bibinfo{author}{\bibfnamefont{Y.~V.} \bibnamefont{Nazarov}},
  \bibinfo{journal}{Phys. Rev. B} \textbf{\bibinfo{volume}{67}},
  \bibinfo{pages}{085316} (\bibinfo{year}{2003}).

\bibitem[{\citenamefont{Shelankov and Rammer}(2003)}]{Shelankov2003}
\bibinfo{author}{\bibfnamefont{A.}~\bibnamefont{Shelankov}} \bibnamefont{and}
  \bibinfo{author}{\bibfnamefont{J.}~\bibnamefont{Rammer}},
  \bibinfo{journal}{Europhys. Lett.} \textbf{\bibinfo{volume}{63}},
  \bibinfo{pages}{485} (\bibinfo{year}{2003}).

\bibitem[{sup()}]{sup}
\emph{\bibinfo{title}{See appendix for more detail.}}

\bibitem[{\citenamefont{M.-Thomson et~al.}(1972)\citenamefont{M.-Thomson,
  Abramowitz, and Stegun}}]{Milne1972}
\bibinfo{author}{\bibfnamefont{L.~M.} \bibnamefont{M.-Thomson}},
  \bibinfo{author}{\bibfnamefont{M.}~\bibnamefont{Abramowitz}},
  \bibnamefont{and} \bibinfo{author}{\bibfnamefont{I.}~\bibnamefont{Stegun}},
  \emph{\bibinfo{title}{Handbook of Mathematical Functions}}
  (\bibinfo{publisher}{Dover Publications}, \bibinfo{year}{1972}).

\bibitem[{\citenamefont{Splettstoesser
  et~al.}(2007)\citenamefont{Splettstoesser, Governale, K{\"o}nig, Taddei, and
  Fazio}}]{Janine2007}
\bibinfo{author}{\bibfnamefont{J.}~\bibnamefont{Splettstoesser}},
  \bibinfo{author}{\bibfnamefont{M.}~\bibnamefont{Governale}},
  \bibinfo{author}{\bibfnamefont{J.}~\bibnamefont{K{\"o}nig}},
  \bibinfo{author}{\bibfnamefont{F.}~\bibnamefont{Taddei}}, \bibnamefont{and}
  \bibinfo{author}{\bibfnamefont{R.}~\bibnamefont{Fazio}},
  \bibinfo{journal}{Phys. Rev. B} \textbf{\bibinfo{volume}{75}},
  \bibinfo{pages}{235302} (\bibinfo{year}{2007}).

\end{thebibliography}

\end{document}